\documentclass[%
reprint,
superscriptaddress,
nofootinbib,
nobibnotes,
 amsmath,amssymb,
 aps,
 prd,
 longbibliography,
 twocolumn,
]{revtex4-2}

\usepackage{float}
\usepackage[percent]{overpic}
\usepackage{graphicx, comment}
\usepackage{dcolumn}
\usepackage{bm}
\usepackage{xcolor}
\usepackage{hyperref}
\usepackage[mathlines]{lineno}
\usepackage[normalem]{ulem}
\usepackage{amsfonts}
\usepackage{url}

\begin{document}

\preprint{APS/123-QED}

\title{First Measurement of Energy-Dependent Inclusive Muon Neutrino Charged-Current Cross Sections 
on Argon with the MicroBooNE Detector}

\newcommand{\Bern}{Universit{\"a}t Bern, Bern CH-3012, Switzerland}
\newcommand{\BNL}{Brookhaven National Laboratory (BNL), Upton, NY, 11973, USA}
\newcommand{\UCSB}{University of California, Santa Barbara, CA, 93106, USA}
\newcommand{\Cambridge}{University of Cambridge, Cambridge CB3 0HE, United Kingdom}
\newcommand{\CIEMAT}{Centro de Investigaciones Energ\'{e}ticas, Medioambientales y Tecnol\'{o}gicas (CIEMAT), Madrid E-28040, Spain}
\newcommand{\Chicago}{University of Chicago, Chicago, IL, 60637, USA}
\newcommand{\Cincinnati}{University of Cincinnati, Cincinnati, OH, 45221, USA}
\newcommand{\CSU}{Colorado State University, Fort Collins, CO, 80523, USA}
\newcommand{\Columbia}{Columbia University, New York, NY, 10027, USA}
\newcommand{\Edinburgh}{University of Edinburgh, Edinburgh EH9 3FD, United Kingdom}
\newcommand{\FNAL}{Fermi National Accelerator Laboratory (FNAL), Batavia, IL 60510, USA}
\newcommand{\Granada}{Universidad de Granada, Granada E-18071, Spain}
\newcommand{\Harvard}{Harvard University, Cambridge, MA 02138, USA}
\newcommand{\IIT}{Illinois Institute of Technology (IIT), Chicago, IL 60616, USA}
\newcommand{\KSU}{Kansas State University (KSU), Manhattan, KS, 66506, USA}
\newcommand{\Lancaster}{Lancaster University, Lancaster LA1 4YW, United Kingdom}
\newcommand{\LANL}{Los Alamos National Laboratory (LANL), Los Alamos, NM, 87545, USA}
\newcommand{\Manchester}{The University of Manchester, Manchester M13 9PL, United Kingdom}
\newcommand{\MIT}{Massachusetts Institute of Technology (MIT), Cambridge, MA, 02139, USA}
\newcommand{\Michigan}{University of Michigan, Ann Arbor, MI, 48109, USA}
\newcommand{\Minnesota}{University of Minnesota, Minneapolis, MN, 55455, USA}
\newcommand{\NMSU}{New Mexico State University (NMSU), Las Cruces, NM, 88003, USA}
\newcommand{\Oxford}{University of Oxford, Oxford OX1 3RH, United Kingdom}
\newcommand{\Pitt}{University of Pittsburgh, Pittsburgh, PA, 15260, USA}
\newcommand{\Rutgers}{Rutgers University, Piscataway, NJ, 08854, USA}
\newcommand{\SLAC}{SLAC National Accelerator Laboratory, Menlo Park, CA, 94025, USA}
\newcommand{\SDSMT}{South Dakota School of Mines and Technology (SDSMT), Rapid City, SD, 57701, USA}
\newcommand{\Maine}{University of Southern Maine, Portland, ME, 04104, USA}
\newcommand{\Syracuse}{Syracuse University, Syracuse, NY, 13244, USA}
\newcommand{\TelAviv}{Tel Aviv University, Tel Aviv, Israel, 69978}
\newcommand{\Tennessee}{University of Tennessee, Knoxville, TN, 37996, USA}
\newcommand{\UTA}{University of Texas, Arlington, TX, 76019, USA}
\newcommand{\Tufts}{Tufts University, Medford, MA, 02155, USA}
\newcommand{\VTech}{Center for Neutrino Physics, Virginia Tech, Blacksburg, VA, 24061, USA}
\newcommand{\Warwick}{University of Warwick, Coventry CV4 7AL, United Kingdom}
\newcommand{\Yale}{Wright Laboratory, Department of Physics, Yale University, New Haven, CT, 06520, USA}

\affiliation{\Bern}
\affiliation{\BNL}
\affiliation{\UCSB}
\affiliation{\Cambridge}
\affiliation{\CIEMAT}
\affiliation{\Chicago}
\affiliation{\Cincinnati}
\affiliation{\CSU}
\affiliation{\Columbia}
\affiliation{\Edinburgh}
\affiliation{\FNAL}
\affiliation{\Granada}
\affiliation{\Harvard}
\affiliation{\IIT}
\affiliation{\KSU}
\affiliation{\Lancaster}
\affiliation{\LANL}
\affiliation{\Manchester}
\affiliation{\MIT}
\affiliation{\Michigan}
\affiliation{\Minnesota}
\affiliation{\NMSU}
\affiliation{\Oxford}
\affiliation{\Pitt}
\affiliation{\Rutgers}
\affiliation{\SLAC}
\affiliation{\SDSMT}
\affiliation{\Maine}
\affiliation{\Syracuse}
\affiliation{\TelAviv}
\affiliation{\Tennessee}
\affiliation{\UTA}
\affiliation{\Tufts}
\affiliation{\VTech}
\affiliation{\Warwick}
\affiliation{\Yale}

\author{P.~Abratenko} \affiliation{\Tufts} 
\author{R.~An} \affiliation{\IIT}
\author{J.~Anthony} \affiliation{\Cambridge}
\author{L.~Arellano} \affiliation{\Manchester}
\author{J.~Asaadi} \affiliation{\UTA}
\author{A.~Ashkenazi}\affiliation{\TelAviv}
\author{S.~Balasubramanian}\affiliation{\FNAL}
\author{B.~Baller} \affiliation{\FNAL}
\author{C.~Barnes} \affiliation{\Michigan}
\author{G.~Barr} \affiliation{\Oxford}
% ...

\collaboration{The MicroBooNE Collaboration}
\thanks{microboone\_info@fnal.gov}\noaffiliation

\date{\today}

\begin{abstract}
We report a measurement of the energy-dependent total charged-current cross section $\sigma\left(E_\nu\right)$ 
for inclusive muon neutrinos scattering on argon, as well as measurements of flux-averaged differential cross sections as a 
function of muon energy and hadronic energy transfer ($\nu$). Data corresponding to 5.3$\times$10$^{19}$ protons 
on target of exposure were collected using the MicroBooNE liquid argon time projection chamber located in the 
Fermilab Booster Neutrino Beam with a mean neutrino energy of approximately 0.8~GeV. 
The mapping between the true neutrino energy $E_\nu$ and reconstructed neutrino energy $E^{rec}_\nu$ and between 
the energy transfer $\nu$ and reconstructed hadronic energy $E^{rec}_{had}$
are validated by comparing the data and Monte Carlo (MC) predictions. In particular, the modeling of the 
missing hadronic energy and its associated uncertainties are verified by a new method that compares the 
 $E^{rec}_{had}$ distributions between data and an MC prediction after 
constraining the reconstructed muon kinematic distributions, energy and polar angle, to those of data.
The success of this validation gives confidence that the missing energy in the MicroBooNE detector is 
well-modeled and underpins first-time measurements of both the total cross section $\sigma\left(E_\nu\right)$
and the differential cross section $d\sigma/d\nu$ on argon.
\end{abstract}

\maketitle

Current and next-generation precision neutrino oscillation experiments aim to answer several 
critical questions in particle physics~\cite{Diwan:2016gmz} by: i) 
searching for CP violation in the lepton sector~\cite{T2K:2021xwb,NOvA:2021nfi},
ii) determining the neutrino mass ordering~\cite{Qian:2015waa},
and iii) searching for light sterile neutrinos~\cite{Machado:2019oxb}. 
For this purpose, the SBN~\cite{Antonello:2015lea} and DUNE~\cite{Acciarri:2016crz,DUNE:2020lwj}
experiments employ liquid argon time projection chambers 
(LArTPCs)~\cite{rubbia77,Chen:1976pp,willis74,Nygren:1976fe}, a tracking 
calorimeter that enables excellent neutrino flavor identification and 
neutrino energy ($E_{\nu}$) reconstruction in the GeV energy range~\cite{Cavanna:2018yfk}.
These experiments are designed to measure the neutrino flavor oscillations as a function of 
$E_\nu$, which requires a good understanding of the neutrino energy spectrum, 
neutrino-argon interaction cross sections~\cite{Formaggio:2013kya}, and LArTPC 
detector response. High-precision measurements of $\nu$-Ar cross sections, particularly 
those related to energy reconstruction, are of paramount importance. 

While historical accelerator-based neutrino experiments often reported $E_\nu$-dependent cross 
sections~\cite{MINOS:2009ugl, MiniBooNE:2010eis}, recent experiments tend to limit cross-section 
measurements to the directly observable 
lepton and/or hadron kinematics~\cite{Zyla:2020zbs}. This paradigm shift was triggered by concerns 
that quantities not directly measurable in detectors (e.g.~the missing hadronic energy of the interaction 
from undetected neutral particles) may not be correctly modeled in simulations,
which is of particular concern in a broad-band neutrino beam.
In this letter, we demonstrate that the MicroBooNE tune model~\cite{genie-tune-paper} (based on GENIE-v3~\cite{GENIE:2021npt})
of missing energy with its associated uncertainty
can be validated with inclusive muon neutrino charged-current ($\nu_\mu$CC) interactions from the the MicroBooNE 
detector~\cite{Acciarri:2016smi}. After constraining the lepton kinematics distributions of Monte Carlo 
(MC) to those of data, the comparison of reconstructed hadronic energy $E^{rec}_{had}$ distributions between 
data and the updated MC prediction reveals whether the model is able to describe the relationship between 
the lepton kinematics and the visible hadronic energy. This procedure validates whether the 
missing hadronic energy is sufficiently modelled given the prior knowledge of the neutrino flux and 
detector effects. This new procedure enables a first measurement of the differential cross 
section as a function of the energy transfer to the argon $d\sigma/d\nu$.
Together with the differential cross section as a function of the muon 
energy ($d\sigma/dE_\mu$), the $E_\nu$-dependent cross sections are extracted.
These data could be used to isolate problems for low-$E_{\nu}$ cross sections
and to reduce modelling uncertainties for the low-$\nu$ method~\cite{low-nu1,low-nu2,Bodek:2012uu}
to constrain the shape of neutrino energy spectrum in future experiments.


The MicroBooNE detector is a 10.4$\times$2.6$\times$2.3 m$^3$ LArTPC. It consists of approximately 85 ton of liquid Ar 
in the active TPC volume for ionization charge detection, along with 32 
photomultiplier tubes (PMTs)~\cite{pmt} for scintillation light detection.
This work makes use of a data set corresponding to an exposure of 5.3$\times10^{19}$ protons on target (POT)
from the Booster Neutrino Beamline (BNB), which produces a neutrino flux with an estimated 93.6\% $\nu_\mu$
purity~\cite{AguilarArevalo:2008yp} and a mean $E_\nu$ of 0.8~GeV. At these energies, $\nu$-Ar interactions are dominated by
quasielastic and meson-exchange current interactions as well as resonant pion productions, and the final-state hadrons consist
mostly of protons and neutrons with some charged and neutral pions.
The $\mathcal{O}(1)$~MeV energy threshold~\cite{ArgoNeuT:2018tvi} of LArTPC allows detecting
these particles down to low kinetic energies.

Compared to earlier work~\cite{Adams:2019iqc}, this measurement incorporates an improved TPC detector simulation and signal processing
procedure~\cite{Adams:2018dra,Adams:2018gbi}, the Wire-Cell tomographic event reconstruction~\cite{Qian:2018qbv,MicroBooNE:2020vry}, 
and a many-to-many TPC-charge to PMT-light matching algorithm for cosmic-ray rejection~\cite{MicroBooNE:2020vry}.
In particular, the ``generic neutrino detection''~\cite{Abratenko:2020sxa,MicroBooNE:2021zul},
which limits the cosmic-ray muon backgrounds to below 15\% at over 80\% $\nu_\mu$CC selection efficiency,  
is used as a pre-selection. The $\nu_\mu$CC event selection is further improved using a set of pattern recognition techniques,
including i) neutrino vertex identification, ii) track/shower topology separation, iii) particle identification,
and iv) particle flow reconstruction in the Wire-Cell reconstruction package~\cite{wc_pattern_recognition}.
Since many of the analysis details in this work including the event reconstruction, event selection, the overall 
model prediction along with its systematic uncertainties, and the model validation are in common with those in 
searching for an anomalous low-energy excess in inclusive charged-current $\nu_e$ channel that was documented in Ref.~\cite{WC_PRD}, they are only briefly reviewed in this letter.

First, the reconstructed neutrino vertex is required to be inside a fiducial volume,
defined to be 3~cm inside the effective detector boundary~\cite{MicroBooNE:2021zul}. 
Second, a set of dedicated background taggers are constructed to further reject residual
muon backgrounds that entered the detector from outside based on directional information.
Finally, neutral-current (NC) events are substantially reduced by requiring a 
reconstructed primary muon candidate to be longer than 5~cm. 
Some limited charged pion rejection is achieved by detecting large-angle 
scattering in reconstructed track trajectories.
Using input variables from the background taggers, a multivariate classifier is constructed using
the modern boosted decision tree (BDT) library XGBoost~\cite{Chen:2016btl} that yields a $\nu_\mu$CC
selection with an estimated 92\% purity and 68\% efficiency~\cite{WC_PRD}.
In total, 11528 $\nu_\mu$CC candidates are selected and used for cross-section extraction.
About 1/3 of the events are fully contained (FC) and 2/3 are partially contained (PC).
Here, the FC events are defined to be events with their main TPC cluster~\cite{MicroBooNE:2020vry} 
fully contained within the fiducial volume~\cite{MicroBooNE:2021zul} and PC events are mostly 
because of exiting muons. 

\begin{figure}[htp]
  \centering
  \begin{overpic}[width=0.5\textwidth]{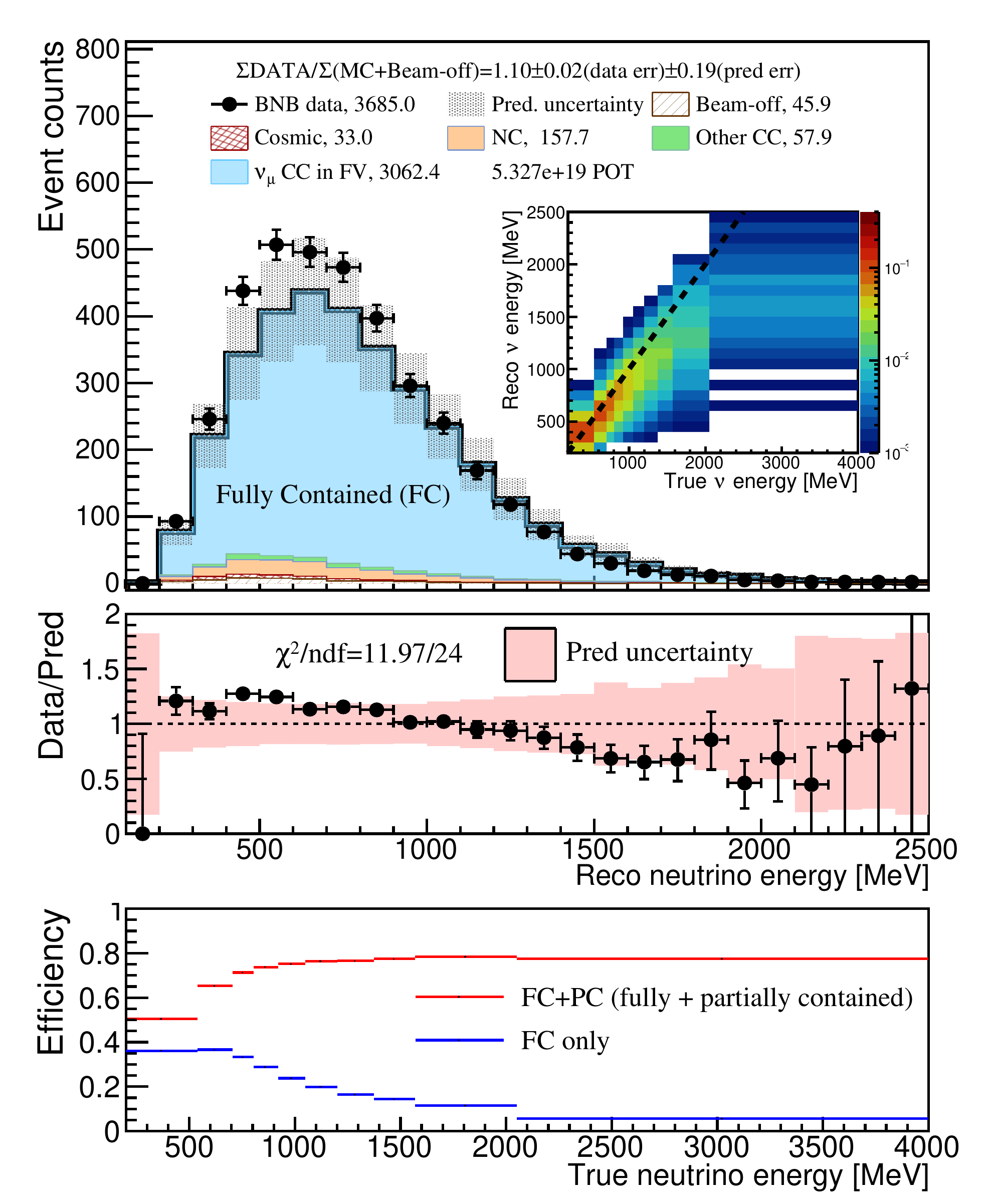}
   \put(11,97){\textsf{MicroBooNE $\mathsf{5.3\times10^{19}}$POT}}
  \end{overpic}
\caption{(Top) Distribution of the selected FC $\nu_{\mu}$CC events as a function of reconstructed neutrino energy. The inset figure is the smearing matrix from true neutrino energy to reconstructed neutrino energy. (Middle) Data/prediction ratio. The pink band represents the total uncertainty (statistical and systematic) of the MC prediction. (Bottom) Selection efficiency of the $\nu_{\mu}$CC events in the fiducial volume as a function of true neutrino energy. At high neutrino energy, muons are more likely to exit the TPC, which leads an 
increase (decrease) in the efficiency of PC (FC) samples.
}
\label{fig:numuCC}
\end{figure}

Three methods are used to reconstruct the energy of tracks and electromagnetic (EM) 
showers~\cite{wc_pattern_recognition, WC_PRD}: i) The energy of a stopped charged particle track 
can be estimated by its travel {\it range} using the NIST PSTAR database~\cite{pstar}. ii) The 
kinetic energy of a charged particle track can be estimated by integrating over the 
reconstructed energy loss per unit length $dE/dx$, which is calculated from 
the measured $dQ/dx$ (ionization charge per unit length) using a {\it recombination} model~\cite{Adams:2019ssg}. 
iii) The energy of an EM shower can be estimated calorimetrically by {\it scaling} the total reconstructed charge of 
the EM shower with a factor of 2.50, which is derived from simulation and includes 
the bias in the reconstructed charge~\cite{MicroBooNE:2020vry} and the average recombination 
factor. This factor is validated with the reconstructed invariant mass of the neutral 
pion~\cite{Adams:2019law}. For stopping tracks with trajectories longer than 4~cm, the {\it range}
method is used to estimate the energy. For short tracks ($<$4~cm), tracks exiting
the detector, tracks with ``wiggled'' topology~\cite{wc_pattern_recognition} (e.g.~low-energy electrons), and muon tracks 
with identified $\delta$ rays, the {\it recombination} method is used to estimate its kinetic 
energy.

The reconstructed neutrino energy $E^{rec}_\nu$ per event is estimated by
summing the kinetic energies of each reconstructed (visible) final-state particle. 
For each reconstructed muon, charged 
pion, or electron candidate, its mass is added to the energy reconstruction.
An average binding energy of 8.6~MeV~\cite{Sukhoruchkin:amnbe} was added for each proton identified.
Figure~\ref{fig:numuCC} shows the FC $\nu_\mu$CC distribution as a function of
reconstructed neutrino energy, the selection efficiency as a function of true
neutrino energy, and the smearing matrix between $E^{rec}_\nu$ and $E_\nu$ according to 
the Monte Carlo simulation.
The predicted energy resolution using the MicroBooNE MC for FC $\nu_\mu$CC events is $\sim$10\% for muon energy, 
$\sim$20\% for neutrino energy, and $\sim$30-50\% for hadronic energy.
The hadronic energy resolution is dominated by the missing hadronic energy and imperfect event 
reconstruction. For events well reconstructed,
the resolution of the reconstructed visible hadronic energy approaches $\sim$10\%.
Among all events, the average bias (towards low energy) of $E^{rec}_\nu$ for FC $\nu_\mu$CC events 
is less than 10\% for $E_\nu<$~800~MeV and increases to $\sim$25\% at $E_\nu=$~2.5~GeV.

The total and differential cross sections are extracted using the Wiener-SVD unfolding
method~\cite{Tang:2017rob} as follows:
\begin{equation}\label{eq:master}
    M_i -B_i= \sum_j R_{ij} \cdot S_j = \sum_j \widetilde{\Delta}_{ij} \cdot \widetilde{F}_{j} \cdot S_{j}.
\end{equation}
$M_i$ is the measured number of events in bin $i$ of the reconstructed energy space,
and $B_i$ is the expected number of backgrounds.
$R_{ij} = \widetilde{\Delta}_{ij} \cdot \widetilde{F}_{j}$ is the overall response matrix.
$S_j$, to be extracted, is the average (differential) cross section in bin $j$ of
the true energy, weighted by the nominal $\nu_\mu$ neutrino flux, which is tabulated
in Ref.~\cite{Adams:2019iqc}. This definition of $S_{j}$ with the nominal neutrino flux 
coincides with a recommendation from Ref.~\cite{Koch:2020oyf} in addressing a 
concern on the treatment of neutrino flux uncertainty.
$\widetilde{\Delta}_{ij}$, the ratio between the selected number of events in reconstructed
energy bin $i$ that originate from the true energy bin $j$ and the generated number of events
in bin $j$, is calculated using central-value MC. This encapsulates both the
smearing between reconstructed and true space and the efficiency. $\widetilde{F}_{j}$ is a constant 
that is calculated with the POT, number of Ar nuclei, the integrated nominal $\nu_\mu$ 
flux in bin $j$, and the bin width (for differential cross sections only). 


The Wiener-SVD unfolding is performed based on a 
\begin{equation}
    \chi^2 = \left(\bm{M}-\bm{B}-\bm{R}\cdot \bm{S} \right)^T\cdot \bm{V}^{-1} \cdot \left(\bm{M}-\bm{B} -\bm{R}\cdot \bm{S}\right)
\end{equation}
test statistics and an additional regularization constructed from a Wiener filter~\cite{Tang:2017rob}.
$\bm{V}$ is the covariance matrix on the measured number of events
in the reconstructed energy bins, encoding the statistical and systematic uncertainties for both signal and background events.
Statistical uncertainties on the data are calculated following the combined Neyman-Pearson procedure~\cite{Ji:2019yca}. 

The covariance matrix also includes several systematic uncertainties. The neutrino flux model
uncertainty (5-15\%) follows the work in Ref.~\cite{Adams:2019iqc}. It includes effects from
hadron production of $\pi^{+}$, $\pi^{-}$, $K^{+}$, $K^{-}$, and $K^{0}_{L}$, together with total, inelastic, and quasielastic cross sections of pion and nucleon 
re-scattering on beryllium and aluminum. In addition, modeling of the
horn current distribution and calibration is included. 
The neutrino-argon interaction cross section model uncertainties  ($\sim$20\%) are described in Ref.~\cite{genie-tune-paper}.
Particularly, the uncertainties associated with the hadronic interactions, which are important in modeling missing energy, are conservatively
estimated: the proton to neutron charge exchange and the proton knockout have 50\% and 20\%
uncertainties, respectively~\cite{Andreopoulos:2009rq, Andreopoulos:2015wxa}. 
The uncertainties on the GEANT4 models~\cite{Allison:2016lfl} used to simulate secondary 
interactions of protons and charged pions outside the target nucleus ($\sim$1.5\%) follows Ref.~\cite{Calcutt:2021zck}.
These uncertainties on the flux, cross section, and GEANT4 models are estimated
using a \emph{multisim} technique~\cite{roe2007statistical}
in which parameters that govern interaction models are simultaneously varied in generating
hundreds of universes to construct covariance matrices.

The detector response uncertainty follows the work
in Ref.~\cite{MicroBooNE:2021roa}, considering the effects of variations in the TPC waveform, light yield and propagation, space charge effect~\cite{Abratenko:2020bbx,Adams:2019qrr}, 
and ionization recombination model. For each source, the same set of MC interactions are 
re-simulated through the detector response simulation with a $1\sigma$ change to the 
corresponding detector model parameter. The differences in the selected number of 
events between the modified and original simulations are used to construct a 
covariance matrix with a bootstrapping~\cite{EfroTibs93} procedure.
The uncertainty of modeling the ``dirt'' events that originate outside the cryostat
follows the work in Ref.~\cite{WC_PRD}. The statistical uncertainty of the Monte-Carlo sample is treated
using the methods described in Ref.~\cite{Arguelles:2019izp}. The uncertainties on the POT (2\% based on \emph{in-situ}
proton flux measurements~\cite{AguilarArevalo:2008yp}) and the number of target nuclei ($\sim$1\%) are also included.


Given Eq.~\eqref{eq:master}, the uncertainties on the neutrino flux, GEANT4 model, detector model, and POT 
enter through $B_i$ and the numerator of $\widetilde{\Delta}_{ij}$.
``Dirt'' uncertainties enter through $B_i$. 
In comparison, the cross section uncertainty enters through $B_i$ and both numerator and denominator of $\widetilde{\Delta}_{ij}$.
Although the uncertainty on the predicted inclusive cross section is $\sim$20\%, it is reduced to $\sim$5\% 
because of the cancellation between numerator and denominator of $\widetilde{\Delta}_{ij}$.


\begin{figure}[htp]
  \centering
  \begin{overpic}[width=0.48\textwidth]{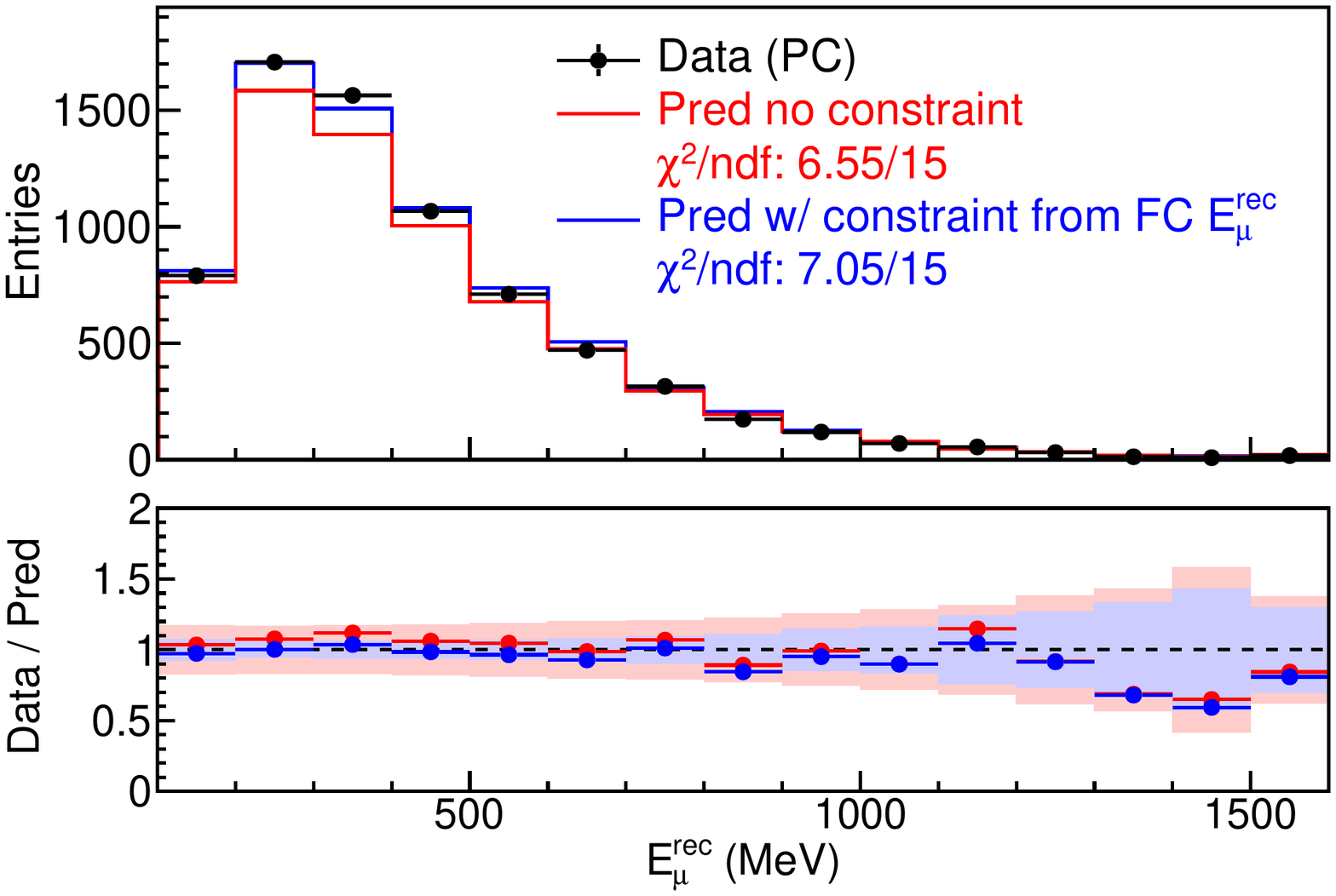}
   \put(43,4){\footnotesize{a)}}
   \put(12,68){\textsf{MicroBooNE $\mathsf{5.3\times10^{19}}$POT}}
  \end{overpic}
  
  
  \begin{overpic}[width=0.48\textwidth]{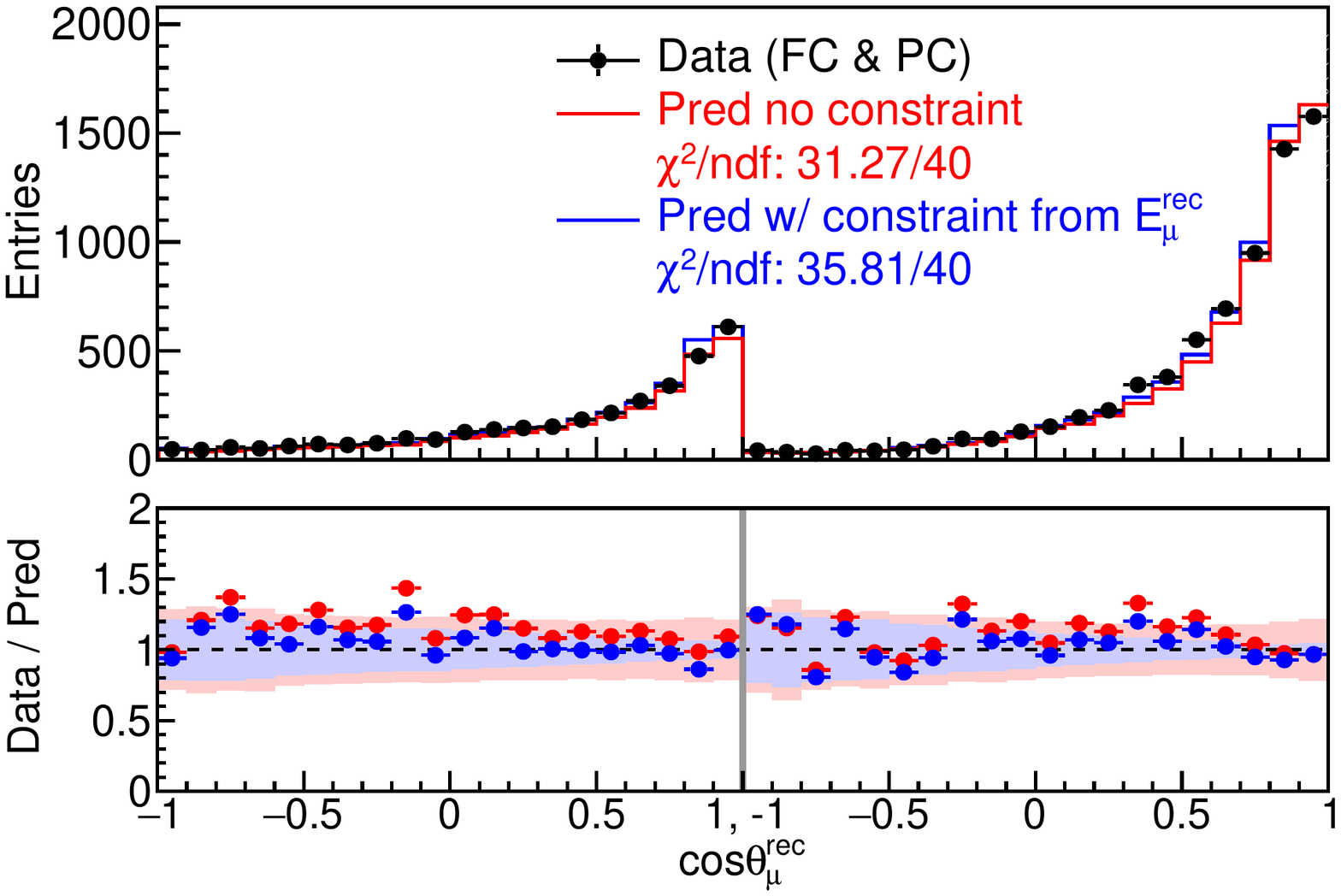}
   \put(44,3.5){\footnotesize{b)}}
  \put(50,36){\footnotesize{\textsf{FC}}}
  \put(92,36){\footnotesize{\textsf{PC}}}
   \put(12,68){\textsf{MicroBooNE $\mathsf{5.3\times10^{19}}$POT}}
  \end{overpic}
  
  
  \begin{overpic}[width=0.48\textwidth]{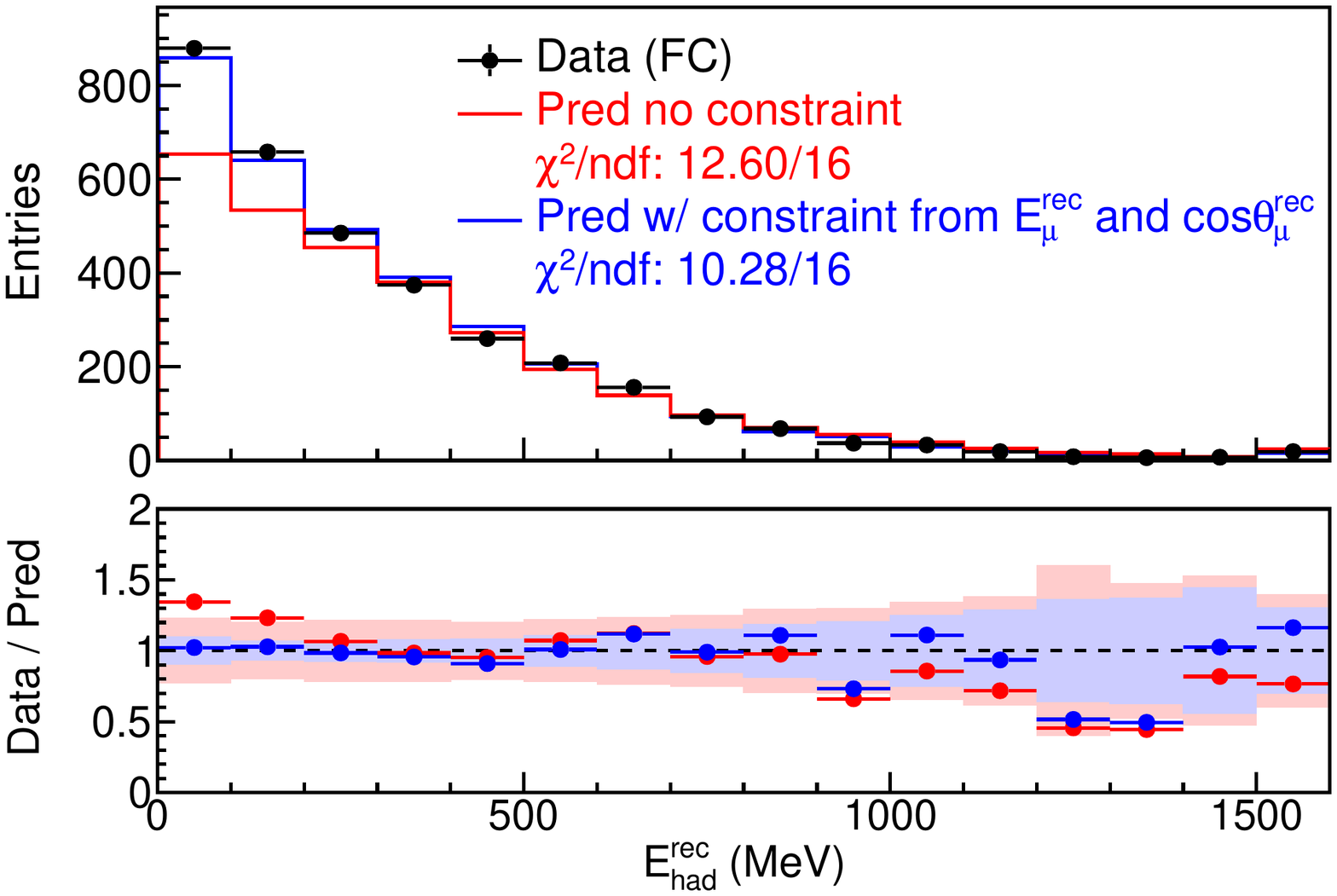}
   \put(43,4){\footnotesize{c)}}
   \put(12,68){\textsf{MicroBooNE $\mathsf{5.3\times10^{19}}$POT}}
  \end{overpic}
\caption{Data are compared with MC predictions as a function of: a) reconstructed 
muon energy $E^{rec}_\mu$ for the partially contained (PC) sample. The MC prediction after 
applying constraints from the fully-contained (FC) sample in $E^{rec}_\mu$ is shown. 
The last bin represents all events with $E^{rec}_\mu>1.5$~GeV. 
The blue (red) points represent the ratio between data and the MC prediction with (without)
constraint, and the bands with same colors depict the total (include statistical) 
uncertainty of the MC prediction.
b) reconstructed $\cos\theta^{rec}_\mu$ for the FC
(first half) and PC (second half) sample. The MC prediction after
applying constraints on both FC and PC samples in $E^{rec}_\mu$
is shown.
c) reconstructed hadronic energy $E^{rec}_{had}$ for the FC sample. The
MC prediction after applying constraints on muon kinematics ($E^{rec}_\mu$ and 
$\cos\theta^{rec}_\mu$) is shown. The last bin represents all events with 
$E^{rec}_{had}>1.5$~GeV.} 
\label{fig:validations}
\end{figure}

A prior condition of using the Wiener-SVD unfolding method to extract cross sections is that the data 
must be well-described by the overall model prediction within its uncertainties. 
In Fig.~\ref{fig:numuCC} and Fig.~\ref{fig:validations}, data and simulation are shown for
key reconstructed kinematic
variables including i) neutrino energy $E^{rec}_{\nu}$, ii) muon energy $E^{rec}_\mu$, iii) cosine
of muon polar angle $\cos\theta^{rec}_{\mu}$, and iv) hadronic energy $E^{rec}_{had}$.
The compatibility between the data and prediction is
demonstrated quantitatively by decent $\chi^2/\mathrm {ndf}$ values (ndf is the number
of degrees of freedom) with corresponding p-values larger than 0.05
considering full uncertainties using the Pearson $\chi^2$~\cite{comp_teststat}. 
To examine different components of systematic uncertainties, 
we further utilize the conditional covariance matrix formalism~\cite{cond_cov} to
adjust the model prediction and reduce its uncertainties by applying constraints from data.
Figure~\ref{fig:validations}a) shows the comparison of the $E^{rec}_\mu$ distribution for PC $\nu_\mu$CC in data to that of the model prediction after applying constraints from the FC $E^{rec}_\mu$ 
events. While the uncertainties are largely reduced, there is only a small change to $\chi^2/\mathrm {ndf}$.
The data and constrained model agree within uncertainties,
verifying the modeling of the invisible energy of muons outside the active detector 
volume for PC events.
Figure~\ref{fig:validations}b) shows the comparison of the $\cos\theta^{rec}_\mu$ distribution for both FC and PC 
$\nu_\mu$CC candidates in data with the model prediction after applying a constraint from the $E^{rec}_\mu$ 
distributions of the same set of $\nu_\mu$CC candidate events. Compared to the previous case, the correlated statistical uncertainties between the $\cos\theta^{rec}_\mu$
distributions and the $E^{rec}_\mu$ distributions are estimated with a bootstrapping procedure.
While the uncertainties are significantly reduced after applying the constraint, the change to
$\chi^2/\mathrm {ndf}$ is small,
showing well-modeled muon kinematics.

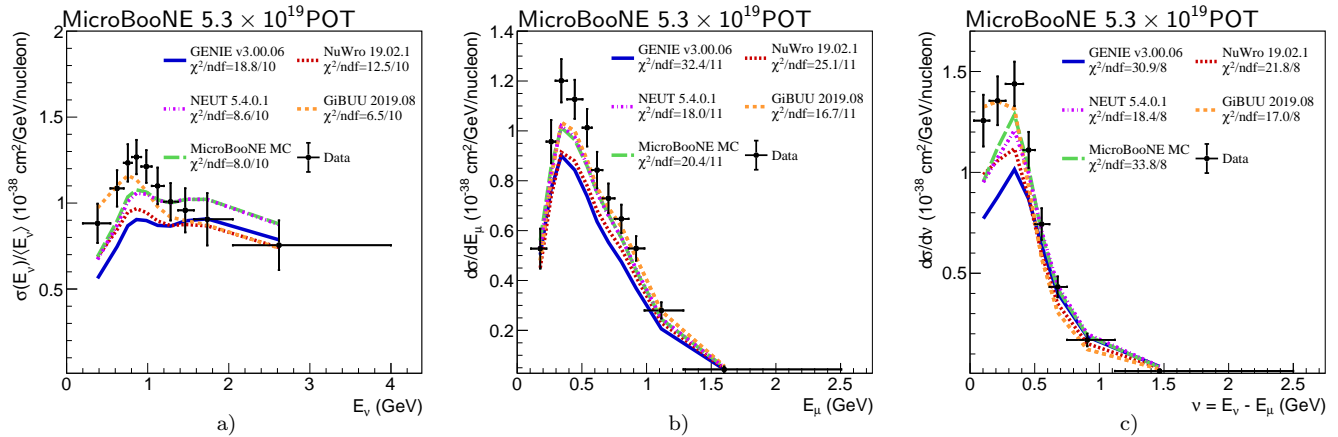
\begin{figure*}[t]
    \centering
    \begin{overpic}[width=0.329\textwidth]{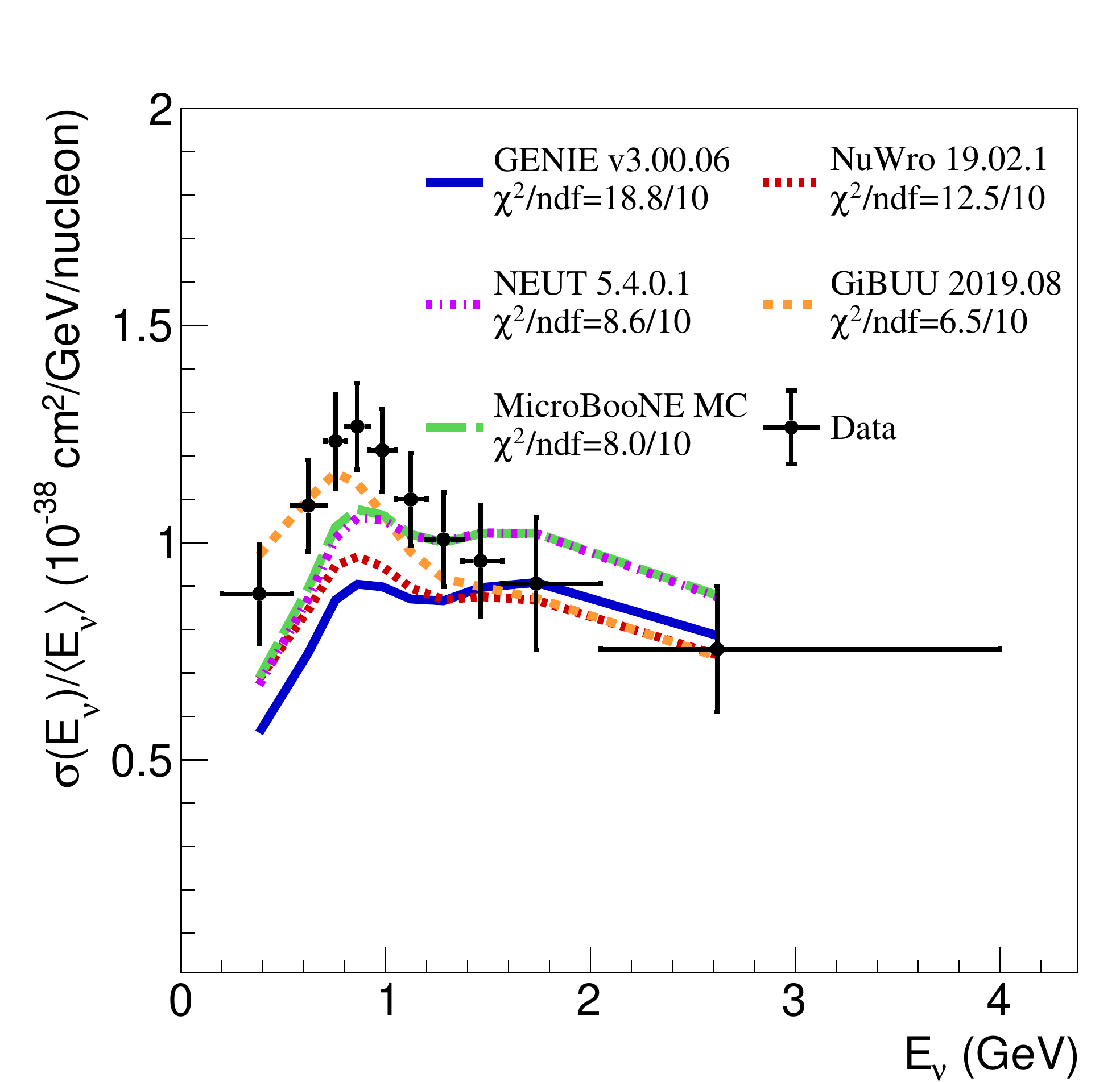}
    \put(50,-3){\footnotesize{a)}}
    \put(15,88){\textsf{MicroBooNE $\mathsf{5.3\times10^{19}}$POT}}
   \end{overpic}
   \begin{overpic}[width=0.329\textwidth]{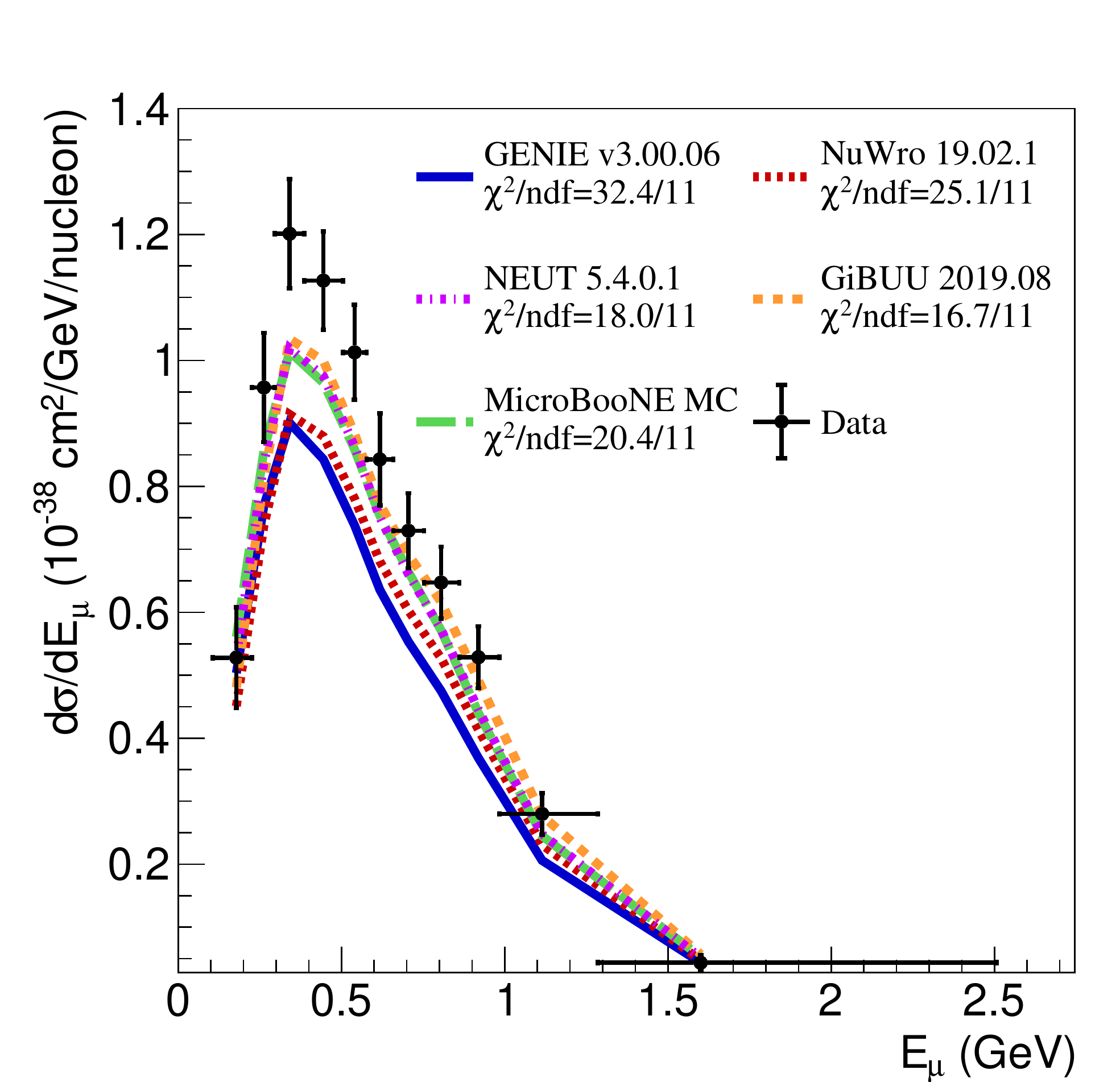}
    \put(50,-3){\footnotesize{b)}}
    \put(16,88){\textsf{MicroBooNE $\mathsf{5.3\times10^{19}}$POT}}
   \end{overpic}
   \begin{overpic}[width=0.329\textwidth]{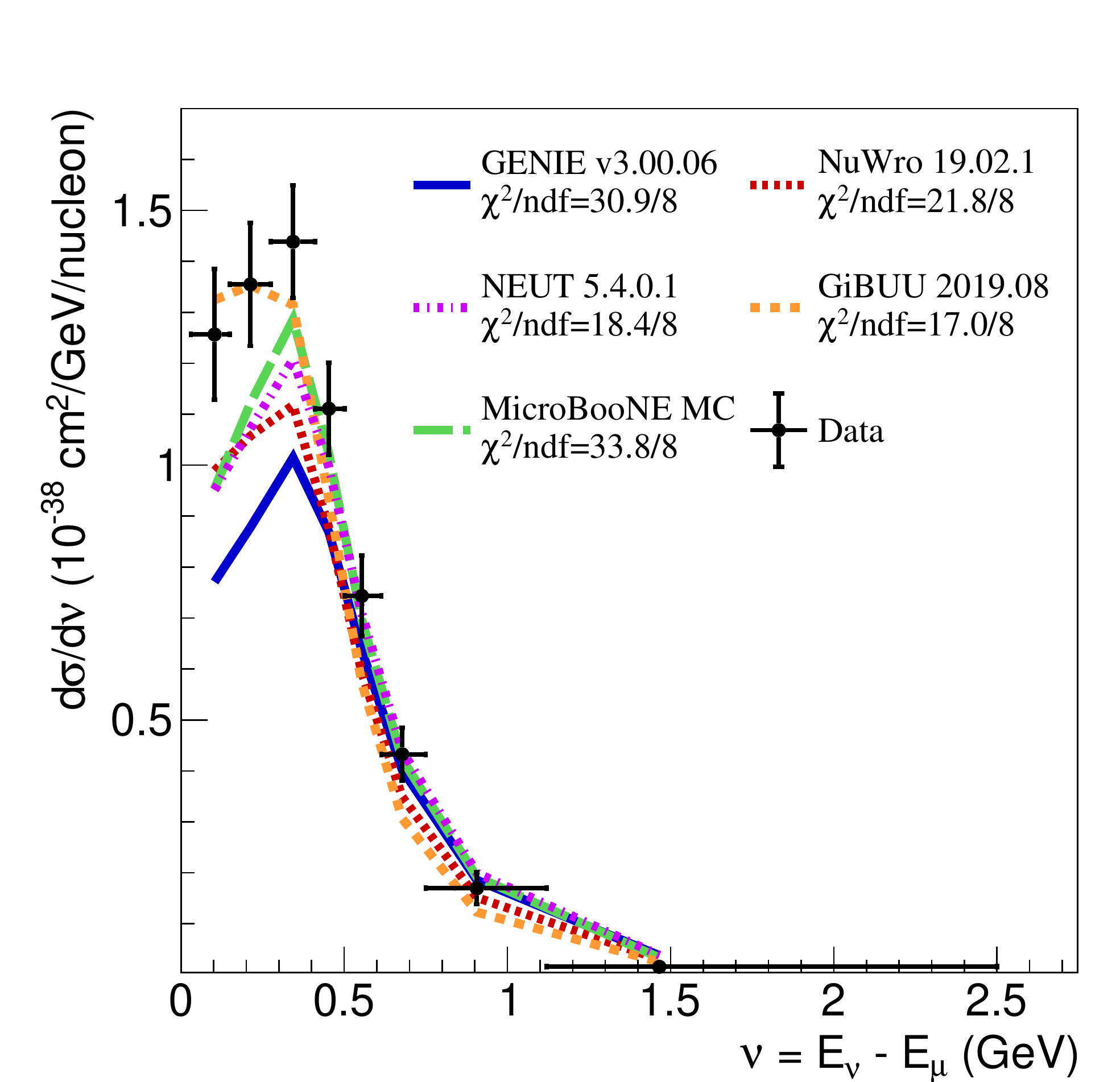}
    \put(50,-3){\footnotesize{c)}}
    \put(15,88){\textsf{MicroBooNE $\mathsf{5.3\times10^{19}}$POT}}
   \end{overpic}
    \caption{a) The extracted $\nu_{\mu}$CC inclusive scattering cross
    section per nucleon divided by the bin-center neutrino energy,
    as a function of neutrino energy. 
    b) The measured $\nu_{\mu}$CC differential cross section per nucleon 
    as a function of muon energy $d\sigma/dE_\mu$. 
    c) The measured $\nu_{\mu}$CC differential cross section per nucleon 
    as a function of energy transfer $d\sigma/d\nu$. Various model predictions are compared to all three measurements (see text for details).}
    \label{fig:xsec_results}
\end{figure*}
We will examine the modeling of the mapping between the reconstructed energy of the hadronic 
system $E^{rec}_{had}$ and the energy transfer to the argon nucleus $\nu = E_\nu - E_\mu$ after taking 
into account the muon results. 
The mapping of $E^{rec}_{had}$ to $\nu$ (or $E^{rec}_\nu$ to true $E_\nu$) relies on the overall cross 
section model to correct for the missing energy going into undetected neutrons, low-energy photons
and other particles below the detection threshold. 
To validate the model, we examine the $E^{rec}_{had}$ distribution for the FC 
$\nu_\mu$CC candidates in data with that of the model prediction after applying constraints from two 
one-dimensional distributions in muon kinematics: $E^{rec}_{\mu}$ and $\cos\theta^{rec}_{\mu}$
in Fig.~\ref{fig:validations}c). After 
applying constraints, the uncertainties on the model prediction for $E^{rec}_{had}$ are significantly 
reduced because of the cancellation of common systematic uncertainties, such as neutrino flux. 
At the lowest energies, it reduces from 20\% to 5\%. Nevertheless, the $\chi^2/\mathrm {ndf}$ values still yields
p-value above 0.5, indicating that the model describes the relationship between 
$E^{rec}_{had}$ and $E^{rec}_\mu$ well within its uncertainty.
In particular, the difference between the data and the model prediction in the first three bins of $E^{rec}_{had}$
is significantly reduced after applying the constraints. 
This test further validates that the modeling of the missing hadronic energy can describe data within its associated 
uncertainty.
We note the conditional covariance matrix formalism, which 
is used to update the MC predictions and their uncertainties
given the data constraints (more details can be found in Ref.~\cite{WC_PRD}), 
is only used in validating the overall model, and is not used in extracting cross 
sections through the unfolding procedure. With fake data, we show that the $\chi^2/\mathrm {ndf}$
has a significant increase with a shift of $\sim$15\% in the hadronic energy fraction 
allocated to protons (mimicking a variation of the proton-inelastic cross section),
and this procedure is also able 
to distinguish between two GENIE models (see Supplemental Material~\cite{suppl}). In addition, the model 
validation procedure is shown to be much more sensitive to detect an insufficient input model 
compared to the extracted cross sections.

With the overall model validated, the total and differential cross sections per nucleon are extracted. 
The binning of the unfolded results is chosen by considering the energy resolution 
and the number of samples in the true space.
Considering both FC and PC samples, the total cross section divided by the bin-center neutrino energy 
is shown as a function of neutrino energy
in Fig.~\ref{fig:xsec_results}a), where the bin-center is calculated as the flux-weighted average neutrino
energy. Excluding the PC sample does not change the overall behaviour of the cross sections, 
but increases their uncertainties for neutrino energy above 1.2~GeV modestly.
Besides the nominal cross-section model used in the ``MicroBooNE MC''~\cite{genie-tune-paper}, 
predictions from GENIE v3.0.6
~\cite{Andreopoulos:2009rq,GENIE:2021npt},
NuWro 19.02.01~\cite{Golan:2012rfa}, NEUT 5.4.0.1~\cite{Hayato:2009zz}, and GiBUU 2019.08~\cite{Buss:2011mx}
after applying the Wiener filter are quantitatively 
compared with the measurement through calculating $\chi^2/{\rm ndf}$ with the uncertainty
covariance matrix obtained from the unfolding procedure. 
Note that these comparisons only incorporate the central predictions from various generators 
without their theoretical uncertainties, which are particularly important in 
constructing predictions in analysis. 
The central predictions of GENIE v3 and NuWro are disfavored compared to the other three. 
Particularly, the ``MicroBooNE MC'' (tuned GENIE-v3 model~\cite{genie-tune-paper}) has better 
agreement than GENIE v3.0.6, given the tuned GENIE-v3 model is constructed by fitting T2K data~\cite{T2K:2016jor} 
in a similar energy range. 

Figure~\ref{fig:xsec_results}b) and c) show the
flux-averaged differential cross sections as a function of muon energy ($d\sigma/dE_\mu$)
and energy transfer to the argon nucleus ($d\sigma/d\nu$).
The same set of model predictions are compared to these measurements. 
The model comparison of $d\sigma/dE_\mu$ shows a shape agreement with most models, although the
normalization predictions differ. The central predictions of GENIE v3 and NuWro are more disfavored.
The model predictions in $d\sigma/d\nu$ show large variations, particularly in the low energy transfer 
($\nu$) region, where the shape difference contributes considerably to the $\chi^2/{\rm ndf}$ 
given the correlations in the uncertainty covariance matrix.
The central prediction from GiBUU has the best agreement with data in the low $\nu$ region, 
but is systematically lower than data at high $\nu$ region, which could be originated
from an underestimation of the cross sections in the nucleon resonance region beyond $\Delta$.
Considering all three cross-section results, the GiBUU prediction has the best agreement with 
acceptable $\chi^2/{\rm ndf}$ values, while the performance of the NEUT prediction is comparable. 
The central predictions of the other three models show larger disagreement.


In summary, we present a measurement of cross section as a function of the neutrino energy
based on data from a broad-band neutrino beam. We report the nominal-flux weighted total inclusive 
$\nu_\mu$CC cross sections $\sigma\left(E_\nu\right)$, and 
the nominal flux-averaged differential cross sections as a function of muon energy
$d\sigma/dE_\mu$ and energy transfer 
$d\sigma/d\nu$ using the Wiener-SVD unfolding method~\cite{Tang:2017rob}. 
A new procedure based on the conditional covariance matrix formalism~\cite{cond_cov} and the 
bootstrapping method~\cite{EfroTibs93} is used to validate the model of missing energies, 
which enables the first measurement of $d\sigma/d\nu$ on argon and 
significantly adds value to the measurement of the total cross section as 
function of neutrino energy $\sigma\left(E_\nu\right)$. 
These results provide a detailed way to compare data 
and calculations beyond what is possible with existing flux-averaged total cross section results. 
With additional accumulated data statistics (up to 1.2$\times10^{21}$ POT from BNB) in the MicroBooNE detector, 
additional neutrino cross-section measurements are expected that will lead to further model development and 
generator improvements for neutrino scattering in argon. 

\bigskip

\begin{acknowledgments}
This document was prepared by the MicroBooNE collaboration using the resources of the Fermi National Accelerator Laboratory (Fermilab), a U.S. Department of Energy, Office of Science, HEP User Facility. Fermilab is managed by Fermi Research Alliance, LLC (FRA), acting under Contract No.~DE-AC02-07CH11359.  MicroBooNE is supported by the following: the U.S. Department of Energy, Office of Science, Offices of High Energy Physics and Nuclear Physics; the U.S. National Science Foundation; the Swiss National Science Foundation; the Science and Technology Facilities Council (STFC), part of the United Kingdom Research and Innovation; the Royal Society (United Kingdom); and The European Union's Horizon 2020 Marie Sklodowska-Curie Actions. Additional support for the laser calibration system and cosmic ray tagger was provided by the Albert Einstein Center for Fundamental Physics, Bern, Switzerland. We also acknowledge the contributions of technical and scientific staff to the design, construction, and operation of the MicroBooNE detector as well as the contributions of past collaborators to the development of MicroBooNE analyses, without whom this work would not have been possible.
\end{acknowledgments}

\bibliography{xs_wc_prl}

\end{document}